\documentclass[english]{lni}

\IfFileExists{latin1.sty}{\usepackage{latin1}}{\usepackage{isolatin1}}

\usepackage{graphicx}
\usepackage{fancyvrb}
\usepackage{amssymb}
\usepackage{amsmath}
\usepackage{array}
\usepackage[table]{xcolor}
\usepackage{relcode, refcount}

\newcommand{\KeY}{Ke\kern-0.1emY~}

\newcommand{\code}[1]{\texttt{#1}}

\author{
Aboubakr Achraf El Ghazi, Ulrich Geilmann, Mattias Ulbrich, Mana Taghdiri \\\\
Karlsruhe Institute of Technology, Germany\\
\{elghazi, geilmann, mulbrich, taghdiri\}@ira.uka.de
}
\title{A Dual-Engine for Early Analysis of Critical Systems}
\begin{document}
\maketitle

\begin{abstract}
This paper presents a framework for modeling, simulating, and checking properties of critical systems based on the Alloy language -- a declarative, first-order, relational logic with a built-in transitive closure operator. The paper introduces a new dual-analysis engine that is capable of providing both \emph{counterexamples} and \emph{proofs}. Counterexamples are found fully automatically using an SMT solver, which provides a better support for numerical expressions than the existing Alloy Analyzer. Proofs, however, cannot always be found automatically since the Alloy language is undecidable. Our engine offers an economical approach by first trying to prove properties using a fully-automatic, SMT-based analysis, and switches to an interactive theorem prover only if the first attempt fails. This paper also reports on applying our framework to Microsoft's COM standard and the mark-and-sweep garbage collection algorithm.
\end{abstract}


\section{Introduction}

Critical infrastructures such as E-Traffic, E-Energy, and Cloud employ various protocols to ensure self-organization, self-reconfiguration, load distribution, and failure recovery. Due to the size, heterogeneousness, and the highly-dynamic nature of those infrastructures, their protocols are often complex, and thus it is crucial to check their security and functionality requirements not only after they are implemented and deployed, but also at their early stages of algorithm design and refinement. This ensures that certain mistakes are caught early, and thus can be fixed at a lower cost.

Lightweight formal methods \cite{lightweight} provide a promising framework for checking critical software systems continuously in earlier stages. Alloy \cite{alloy-book}, for example, provides an expressive, declarative language that can be analyzed fully automatically. The language is a combination of first-order logic and relational algebra, augmented with a built-in transitive closure operator which makes it particularly suitable for modeling structure-rich systems such as network protocols. 

Alloy has been used for checking security and functionality aspects of several resource management, network communication, transportation, and security protocols, supporting the contention that lightweight formal methods are feasible and economical for critical systems. Case studies include a role-based access control security schema for protecting the access to sensitive information and resources~\cite{zao-SACMAT2003}, the intentional naming system for resource discovery in dynamic networked environments~\cite{Khurshid-ASE2000}, a pull-based asynchronous rekeying framework for scalable management of group keys in secure multicast~\cite{taghdiri-FORTE2003}, the NASA's Direct-To system for helping air traffic controllers find flight plans that safely shorten the flying time~\cite{mohsen-mit2001}, the New York City subway signaling system \cite{sarma-ics2001}, the flash file system that caused the famous 18-day system breakdown of the NASA's mars rover \emph{Spirit}~\cite{kang-abz2008}, a constraint analysis on Java Bytecodes to detect security vulnerabilities~\cite{reynolds-ABZ2010}, the security domain model analysis to identify illicit information flows and covert channel vulnerabilities~\cite{shaffer-seke2008}, and the Mondex electronic purse system for decentralized electronic money transactions~\cite{ramananandro_mondex_2007}. 

While all the above case studies use Alloy to check protocols at the design level, several tools have been developed that use Alloy for code-level software checking. Jalloy~\cite{jalloy}, JForge~\cite{forge}, and Karun~\cite{taghdiri-ase-journal}, for example, check functional properties of Java programs via translation to Alloy. TestEra~\cite{testera}, on the other hand, uses Alloy for systematic test case generation.

There are three main reasons for Alloy's popularity: (1) expressiveness of the language, (2) its fully automatic analysis engine, and (3) support for various abstraction levels. Unlike typical model checkers that only check temporal safety properties specified as finite state machines, Alloy is particularly suitable for modeling rich properties of structure-intensive systems. Such systems can be expressed in the Z specification language \cite{z-book} as well, but there is little tool support for automatic analysis of Z specifications. On the other hand, domain-specific tools such as AVISPA~\cite{avispa} and Scyther~\cite{scyther} are fully automatic, but they are specially designed to check security protocols and are not suitable for checking general functionality requirements. Furthermore, Alloy's support of various abstraction levels (chosen by the user), from the algorithm design to the actual code specification, makes it possible to check the abstraction refinement properties in a uniform framework. 

Despite all the successful applications of Alloy to critical systems, the Alloy engine lacks certain capabilities essential for checking critical infrastructures. The Alloy Analyzer (AA) analyzes Alloy specifications fully automatically. This analysis, however, is performed with respect to a finite {\em scope} -- a user-provided bound on the size of the analyzed system -- and thus is called \emph{bounded verification}. For critical infrastructures, however, it is essential to have a complete proof of correctness. AA's lack of proof capability results from the fact that it translates Alloy specifications to (satisfiability-equivalent) propositional formulas, and uses a SAT solver to solve those formulas. Consequently, AA provides a poor support for integer arithmetic (handles them with respect to only a small bitwidth), which is essential in modeling smart meters, E-Energy, and E-traffic infrastructures. 

Furthermore, AA's translation of Alloy to propositional logic is exponential in the scope size, causing AA to run out of memory while translating complex systems in even small scopes. Therefore, the user cannot check the system in a desirable scope by even letting AA run longer (for example overnight).

In a previous paper~\cite{elghazi-taghdiri-fm2011}, we described how proof capability can be added to AA without sacrificing its full automation. However, since the Alloy logic is undecidable, it is not always possible to automatically prove properties of the systems expressed in Alloy. In this paper, we present a dual framework, capable of providing both \emph{proofs} and \emph{counterexamples} based on a 3-step strategy: (1) a fully automatic bounded verification based on SMT (Satisfiability Modulo Theories) that potentially improves on AA's scalability and integer support; (2) a fully automatic proof engine based on SMT and unbounded integers that can be incomplete; and (3) a complete\footnote{Modulo integer arithmetic} but interactive proof engine based on the \KeY interactive theorem prover \cite{key-book}. The framework promises an economical approach for the use of formal methods in the context of critical infrastructures, by requiring user interaction as a last resort -- only if it is really needed.  

There are other approaches that implement a similar
tool chain -- from fully automatic to interactive proving -- for other languages. The HOL-Boogie
approach \cite{Bohme08}, for instance, introduces a multi-phase proof
engine for proof obligations emerging from the VCC compiler. The
majority of obligations can automatically be discharged by an SMT
solver. Only the most sophisticated problems are presented to the user
for interactive proving. Another example is the Why system \cite{boogie11why3} that is used
for software verification. Proof obligations can be discharged either
using automatic provers or by opening them in interactive proof
assistants. Our approach provides a similiar tool chain for the Alloy
language, but adds another step to the chain to quickly inspect models
for potential counterexamples during early design stages.

This paper first gives an overview of our analysis framework, then describes various phases of the framework using an example, and finally reports on our experimental results.


\section{Overall Framework}
To our knowledge, all previous attempts to provide proof capability for the Alloy language were based on interactive theorem provers (ITP). Dynamite \cite{Frias:2007}, for example,  proves properties of Alloy specifications using the PVS theorem prover~\cite{pvs} via a translation to fork algebra. Prioni \cite{arkoudas:2003} integrates the Alloy Analyzer with the Athena theorem prover. In these approaches, proof capability comes at the price of user interaction, regardless of the complexity of the problem. Furthermore, to our knowledge, neither approach handles integer arithmetic expressions allowed in the Alloy language.

Compared to ITP, SMT solvers can efficiently handle a rich combination of decidable theories without sacrificing completeness or full automation. Although adding first-order quantifiers to these theories makes them undecidable, recent SMT solving approaches \cite{ge_complete_2009, bonacina_deciding_2009, ge_solving_2009} have shown significant advances in handling quantifiers. Our framework exploits this. In the full-verification mode, it always tries fully automatic SMT solving first, and switches to ITP only if SMT solving fails.

Since trying to prove an invalid property is particularly costly (an SMT solver may output \emph{unknown} or time out, and an ITP may never terminate), our framework starts in the bounded-verification mode, trying to find a counterexample in a finite scope first. This allows the user to increase the scope arbitrarily in order to gain more confidence about the correctness of the property before switching to the full-verification mode. It should be noted that under certain circumstances, a minimum scope can be computed so that correctness for that scope implies already correctness for any scope \cite{momtahan-entcs2005}.

\begin{figure}[htb]
  \begin{center}
    \includegraphics[width=\textwidth]{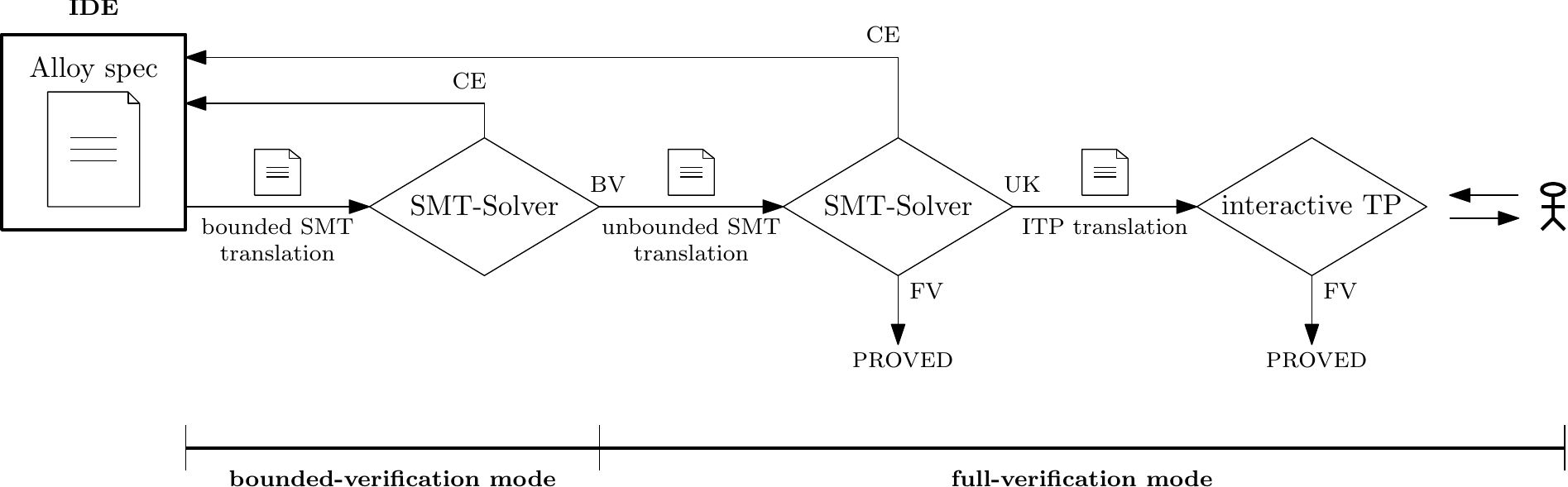}\\
    \vspace*{2mm}
    \caption{Stages of our analysis -- \textbf{CE:} counterexample, \textbf{BV:} bounded valid, \textbf{FV:} fully valid, \textbf{UK:} unknown}
    \label{fig:framework}
  \end{center}
\end{figure}

Figure \ref{fig:framework} gives an overview of our framework. It uses the Alloy IDE to take advantage of Alloy's facilities such as type checking and instance visualization. Technical details of the strategies are discussed in the next section. 



\section{Approach}

Our framework provides three strategies for checking a property of an Alloy specification: (1) {\em Bounded verification} checks Alloy specifications with respect to a bounded scope, aiming at finding counterexamples. Any counterexample reported by this phase is guaranteed to be valid; no false alarms are generated. Lack of a counterexample, however, does not constitute proof; it only implies that no counterexample exists within the analyzed scope. (2) {\em SMT-based full verification} aims at proving the correctness of the property fully automatically using the Z3 SMT solver \cite{z3}. If Z3 outputs ``unsat", the property has been proven correct, and if it outputs a counterexample preceded by the keyword ``sat", a valid counterexample has been found. However, since Alloy is undecidable, Z3 does not guarantee a complete analysis: it may output a counterexample preceded by the keyword ``unknown", implying that the property may or may not be valid, or time out. (3) {\em ITP-based full verification} provides a complete proof engine based on the \KeY theorem prover \cite{key-book}. Due to our extensive set of axioms and lemmas, in some cases, the property can be proved automatically. In general, however, this analysis requires user interactions to guide the theorem prover and thus, its performance depends on the user's level of expertise.

This section describes the basics of the Alloy language and our three analysis strategies using a running example. It focuses on the main ideas involved in each analysis in order to clarify their differences. Technical details and further evaluations of the SMT-based full verification can be found elsewhere \cite{elghazi-taghdiri-fm2011}. Details of our bounded verification and ITP-based full verification will follow in our future publications. 

\subsection{Example}

\begin{figure}
\small{
\begin{relbox}
\nl \open util/ordering[Book] \as ord 										\label{nl:open}
\nl \abstract \sig Target \{\}												\label{nl:target}
\nl \sig Name \extends Target \{\}										\label{nl:name}
\nl \sig Address \extends Target \{\}										\label{nl:addr}
\nl \sig Book \{														\label{nl:book}
\nl 	names: \set Name,												\label{nl:book:names}
\nl 	addr: names \-> \some Target										\label{nl:book:addr}
\nl \}.																\label{nl:book:end}
\nl \fact acyclicity \{ \all* b: Book, n: Name \| \not (n \in n.\^(b.addr)) \}				\label{nl:acyclic}
\nl \fun lookup [b: Book, n: Name]: \set Address \{ n.\^(b.addr) \verb|&| Address \} 	\label{nl:lookup}
\nl \pred add [b, b': Book, n: Name, t: Target] \{								\label{nl:add}
\nl 	(t \in Address) \or (t \in Name \and \some lookup [b, t])					\label{nl:add1}
\nl 	b'.addr \= b.addr + n\->t											\label{nl:add2}
\nl \}.																\label{nl:add:end}
\nl \pred del [b, b': Book, n: Name, t: Target] \{								\label{nl:del}
\nl 	(\no b.addr.n) \or (\some n.(b.addr) \- t)	
\nl 	b'.addr \= b.addr - n\->t
\nl \}.																\label{nl:del:end}
\nl \pred traces [] \{ \label{nl:traces:beg}									\label{nl:traces}
\nl 	\no ord/first.addr												\label{nl:traces:init}
\nl 	\all* b': Book - ord/first \| \letting b = ord/prev[b'] \| \(						\label{nl:traces:rest}
\ul		\some n: Name, t: Target \| add[b, b', n, t] \or del[b, b', n, t]\)
\nl \}.																\label{nl:model:end} \label{nl:traces:end}
\end{relbox}
}
\vspace*{-0.5cm}
\caption{Example -- an address book specified in Alloy}
\label{example}
\vspace*{-0.4cm}
\end{figure}

Figure \ref{example} gives a sample model in Alloy. The model specifies the address book of an email client where names are mapped to email addresses \cite{alloy-book}. Line~\ref{nl:open} imports the \code{ordering} module from the Alloy library to order the elements of type \code{Book}. The ordering functions will be used later by the \code{traces} predicate (Lines \ref{nl:traces:beg} - \ref{nl:traces:end}).
In order to allow the use of aliases and group names for email addresses, the model declares a hierarchical type system (Lines \ref{nl:target} - \ref{nl:addr}) where \code{Name} and \code{Address} are subtypes of the basic type \code{Target}. Types are declared using the {\bf sig} keyword and represent sets of elements. The {\bf extends} keyword specifies that the subtypes are disjoint, and the {\bf abstract} keyword denotes that any element of the supertype must belong to one of the extending subtypes.

Lines \ref{nl:book} - \ref{nl:book:end} declare the \code{Book} type. The field \code{names} represents all the names in the address book, and declares a binary relation of type $\code{Book} \rightarrow \code{Name}$. The multiplicity keyword {\bf set} allows each \code{Book} to be mapped to an arbitrary number of \code{Name}s. The \code{addr} field declares a ternary relation of type $\code{Book} \rightarrow \code{Name} \rightarrow \code{Target}$ where only those elements of \code{Name} that are included in the \code{names} relation are allowed. The multiplicity keyword {\bf some} denotes that for every $b: \code{Book}$, and every $n: \code{Name}$ included in \code{b.names}, the pair $(b, n)$ must be mapped to at least one \code{Target}. 

A fact represents a constraint that is assumed to hold. The acyclicity fact (Line~\ref{nl:acyclic}) constrains that no name can appear in its own set of targets directly or indirectly. The operators~\verb|.|~and~\verb|^|~represent relational join and transitive closure, respectively. 

To describe the dynamic behavior of the system, the model defines additional predicates and functions. The \code{lookup} function (Line~\ref{nl:lookup}) returns all the addresses that correspond to a name in a particular book. The \verb|&| operator denotes set intersection. Predicate \code{add} (Lines \ref{nl:add} - \ref{nl:add:end}) specifies the addition operation: a pair $(n, t): \code{Name} \times \code{Target}$ can be added to an address book $b$ if $t$ is an \code{Address} or if it is a \code{Name} that is already mapped to some \code{Address} in $b$ (directly or indirectly). Predicate \code{del} (Lines \ref{nl:del} - \ref{nl:del:end}) specifies the deletion operation: a pair $(n, t): \code{Name} \times \code{Target}$ can be deleted from an address book $b$ if no name is mapped to $n$ or if $n$ is also mapped to targets other than $t$. The operators~\verb|+|,~\verb|-|, and~$\rightarrow$ denote set union, set difference, and Cartesian product, respectively. 

Predicate \code{traces} (Lines \ref{nl:traces} - \ref{nl:traces:end}) specifies that the \code{Book} elements represent a sequence of \code{add} and \code{del} operations: the first book in the ordering is empty and thus contains no addresses (Line~\ref{nl:traces:init}), and any consecutive pair of books are related by either the \code{add} or \code{del} operation (Line~\ref{nl:traces:rest}).
The rest of this section describes how various properties of this address book model can be analyzed by our different analysis strategies.
\newcommand{\myFrame}[1]{\framebox{\parbox{\linewidth}{ #1 }}}

\subsection{SMT-based Bounded Verification}
Given an Alloy specification and a scope -- a bound on the size of each type, we translate the Alloy specification to a satisfiability-equivalent, bounded SMT problem. Compared to Alloy Analyzer that bit-blasts Alloy problems to SAT, our translation preserves the original structure by axiomatizing Alloy constructs as first-order SMT axioms over bounded sorts -- fixed-sized bitvectors. The resulting SMT problem lies within QBVF (quantified bitvector formula) logic, and thus is decidable \cite{wintersteiger-fmcad2010}. 

Top-level types of the Alloy problem are translated to SMT bitvectors, according to the scope information. Since SMT-Lib -- the standard SMT language -- does not support subtypes, we use membership functions to enforce type hierarchies. For an Alloy subtype $S$, we declare a membership function $isS: T[S] \rightarrow Bool$ where $T[S]$ denotes the top-level supertype of $S$\footnote{In certain cases, it is necessary to declare membership functions for top-level types as well.}. This function denotes which elements of $T[S]$ belong to $S$. To enforce multi-level type hierarchies, additional axioms in the form of membership implications are used.

Relations, the central concept in Alloy, are translated to boolean-valued, uninterpreted, membership functions. An Alloy relation $R: A_1 \rightarrow \ldots \rightarrow A_n$ is represented by an SMT function $R^{smt}: (T[A_1] \times \ldots \times T[A_n]) \rightarrow Bool$, and constrained to hold only elements of its admissible type using the axiom $$\forall a_1: T[A_1], \dots, a_n: T[A_n];\  R^{smt}(a_1, \dots, a_n) \Rightarrow isA_1(a_1) \wedge \ldots \wedge isA_n(a_n)$$ 
Furthermore, multiplicity keywords used in relation declarations are translated using auxiliary SMT functions. For example, to represent a relation declaration $R: A_1 \rightarrow \textbf{some}\ A_2$, we declare an additional, uninterpreted SMT function $oneR: T[A_1] \rightarrow T[A_2]$ that nondeterministically maps each element of $T[A_1]$ to exactly one element of $T[A_2]$. The following axiom is then used to ensure that $R^{smt}$ maps each element of $A_1$ to at least one element of $A_2$ (the semantics of \textbf{some}): $$\forall a: T[A_1];\ isA_1(a) \Rightarrow R^{smt}(a, oneR(a))$$


\begin{figure}[t]
\small{
\begin{relbox}
\nl (\textbf{declare-fun} isAddr (\textbf{BitVec[5]}) \textbf{Bool}) 
\nl (\textbf{declare-fun} isName (\textbf{BitVec[5]}) \textbf{Bool}) 
\ul ;; Target is abstract 
\nl (\assert (\textbf{forall} (t \textbf{BitVec[5]}) (\or (isAddr t)(isName t))))
\ul ;; Addr and Name are disjoint
\nl (\textbf{assert} (\textbf{forall} (t \textbf{BitVec[5]}) (\not (\and (isAddr t) (isName t)))))
\ul ;; names: Book \-> set Name 
\nl (\textbf{declare-fun} names (\textbf{BitVec[5]} \textbf{BitVec[5]}) \textbf{Bool}) 
\nl (\textbf{assert} (\textbf{forall} (b \textbf{BitVec[5]}) (t \textbf{BitVec[5]}) (\=> (names b t) (isName t))))
\ul ;; addr: Book \-> names \-> some Target 
\nl (\textbf{declare-fun} addr (\textbf{BitVec[5]} \textbf{BitVec[5]} \textbf{BitVec[5]}) Bool) 
\nl (\textbf{assert} (\textbf{forall} (b \textbf{BitVec[5]}) (t1 \textbf{BitVec[5]}) (t2 \textbf{BitVec[5]}) 
\ul \hspace{2mm}(\=> (addr b t1 t2) (Book.names b t1))))
\ul ;; Multiplicity keyword ``some" 
\nl (\textbf{declare-fun} oneTarget (\textbf{BitVec[5]} \textbf{BitVec[5]}) \textbf{BitVec[5]}) 
\nl (\textbf{assert} (\textbf{forall} (b \textbf{BitVec[5]}) (t1 \textbf{BitVec[5]}) (t2 \textbf{BitVec[5]})
\ul \hspace{2mm}(\=> (names b t1) (addr b t1 (oneTarget b t1))))) 
\end{relbox}
}
\caption{Example -- bounded SMT translation of the address book declaration part}
\label{fig:smt_type_hierarchy}
\end{figure}       

Figure \ref{fig:smt_type_hierarchy} gives the translation of the declaration part of the address book example. The translation assumes a scope of 32, and thus the top-level types, namely \code{Book} and \code{Target}, are represented by bitvectors of width 5, i.e. \code{BitVec[5]}.

In addition to signatures and relations, Alloy formulas involve set and relational operators. Our translation specifies set-based semantics of these operators using first-order axioms over membership functions. Details of this axiomatization can be  found elsewhere \cite{elghazi-taghdiri-fm2011}.

Using this translation, we can check the following property of the address book model.   

\begin{small}
\begin{relbox}
\ul \assert delUndoesAddBuggy \{
\ul 	\all* b, b', b'': Book, n: Name, t: Target \| \(
\ul 		(add[b, b', n, t] \and del[b', b'', n, t]) \=> b.addr = b''.addr \)
\ul \}.
\ul \check delUndoesAddBuggy \for 32 \label{nl:bv:end}
\end{relbox}
\end{small}

The assertion states that if a pair $(n, t)$ is first added to an address book $b$ and then deleted afterwards, the \code{addr} relation of the final book $b''$ is equal to the that of the original book $b$. As the name suggests, this property is invalid. The Z3 SMT solver finds the following counterexample which represents the case where the initial book $b$ already contains the pair to be added. The book $b'$ after adding $(n,t)$ then also contains the pair, and so deleting it results in the empty book $b''$. The property \code{delUndoesAddBuggy} is therefore violated since $b$ and $b''$ differ. 




\vspace*{3mm}

\myFrame{
\vspace*{-3mm}
\begin{small}
  \begin{tabbing}
     \= \hspace*{7cm} \= \hspace*{1.5cm} \= $\implies$ \= $b''.\mathtt{addr}~$ \=\kill%
    \> (\textbf{define} n bv1[5])
      \>  Alloy types are translated to bitvectors.\\
    \> (\textbf{define} t bv0[5])
      \>  Hence, the counterexample defines \\
    \> (\textbf{define} b bv1[5])
      \> bitvector constants for all symbols in the \\
    \> (\textbf{define} b' bv16[5])
      \> model. For instance: $n$ refers to name 1, \\
    \> (\textbf{define} b'' bv0[5])
      \> $b$ to book 1, $b'$ to book 16, etc.\\
    \> (\textbf{define} (addr (x1 (bv 5)) (x2 (bv 5)) (x3 (bv 5))) 
      \\
    \>\hspace{2mm}(\textbf{if} (\textbf{and} (= x1 bv1[5]) (= x2 bv1[5]) (= x3 bv0[5])) \textbf{true}
      \> \> $\implies$ \> $b.\code{addr} $ \> $=\{ (n,t) \}$ \\
    \> \hspace{2mm}(\textbf{if} (\textbf{and} (= x1 bv16[5]) (= x2 bv1[5]) (= x3 bv0[5])) \textbf{true}
      \> \> $\implies$ \> $b'.\code{addr} $ \> $=\{ (n,t) \}$ \\
    \> \hspace{2mm}\textbf{false}))) 
      \> \> $\implies$ \> $b''.\code{addr} $ \> $=\emptyset$
  \end{tabbing}
\vspace*{-3mm}
\end{small}
}
\vspace*{3mm}

Z3 produces this counterexample in 1.41 seconds whereas the Alloy Analyzer requires 58.01 seconds to find a counterexample in this scope. 

\subsection{SMT-based Full Verification}

The second strategy of our framework is to prove correctness of Alloy assertions fully automatically using the Z3 SMT solver again. In contrast to the bounded analysis that produces an SMT problem in the decidable logic of quantified bitvectors, the unbounded analysis uses the AUFLIA logic \footnote{Closed first-order formulas over the theory of linear integer arithmetic and arrays, extended with free sort and uninterpreted function symbols.} that allows quantifiers over free sorts, and thus is undecidable. Consequently, three outcomes are possible: (1) {\em unsat}, implying that the SMT solver has successfully proven the property correct, (2) a counterexample preceded by the keyword {\em sat}, implying that the SMT solver has successfully found a valid counterexample to the property being checked, and (3) a counterexample preceded by the keyword {\em unknown}, implying that the counterexample may or may not be valid, and must be double-checked. An invalid counterexample denotes an inconclusive analysis. 

The translation rules used in this unbounded, SMT-based analysis are very similar to the ones used in the previous section, except for type declarations. While our bounded translation represents top-level types as fixed-sized bitvectors, our unbounded translation represents them as uninterpreted, free sorts in SMT. Furthermore, the unbounded translation exploits the theory of linear integer arithmetic provided by SMT solvers to translate Alloy's integers. Figure \ref{unbounded-example} shows the unbounded translation of the address book type hierarchy. All other axioms must be rewritten over top-level sorts. Further details can be found elsewhere \cite{elghazi-taghdiri-fm2011}.

\begin{figure}[t]
\small{
\begin{relbox}
\nl (\textbf{declare-sort} Target)
\nl (\textbf{declare-sort} Book)
\nl (\textbf{declare-fun} isAddr (Target) \textbf{Bool}) 
\nl (\textbf{declare-fun} isName (Target) \textbf{Bool}) 
\end{relbox}
}
\label{unbounded-example}
\caption{Example -- unbounded SMT translation of the address book type hierarchy}
\end{figure}


To continue with the address book example, one can fix the \code{delUndoesAddBuggy} property based on the feedback from the previous counterexample. The corrected property \code{delUndoesAdd}, shown below, constrains the initial address book to be empty.
 
\begin{small}
\begin{relbox}
\ul \assert delUndoesAdd \{
\ul 	\all* b, b', b'': Book, n: Name, t: Target \| \(
\ul 		(\no n.(b.addr) \and add[b, b', n, t] \and del[b', b'', n, t]) \=> b.addr = b''.addr \)
\ul \}. \label{nl:smtfv:end}
\end{relbox}
\end{small}

Our bounded analysis reports that this assertion has no counterexamples in the scope of 32. 
Therefore, with confidence in the correctness of the property, we use the unbounded analysis to prove the property correct. Z3 proves this property in 0.01 seconds.




\newcommand{\key}{Ke\kern-0.1emY}

\subsection{ITP-based Full Verification} \label{itp}

In the interactive verification stage, the Alloy model is proven using the \KeY
system. For this purpose, the model is translated to {\key}'s typed first-order logic
(FOL), which also provides support for subtyping.

Since Alloy centers around relations, we need a relational first-order theory. We
therefore introduce the top-level types $Relation$ and $Tuple$ for relations and
their elements, respectively. The uninterpreted predicate $in : Tuple \times
Relation$ connects the two types and denotes membership of a tuple in a relation.

To lower the burden of interactively proving a model correct, the translation
should be as transparent as possible; the correspondence between the original
model and its translation should be obvious. To achieve this, we define a FOL
counterpart for each of the Alloy operators. The semantics of the usual set
operations, like union and intersection, can be axiomatized using the membership
predicate $in$. Defining the semantics of relational operators requires
the possibility to access the components of a tuple. For this purpose, we
introduce subtypes of $Tuple$ and $Relation$ to capture arity information:
\begin{gather*}
\mathit{Atom}, \mathit{Tuple2}, \mathit{Tuple3} ,\ldots <: \mathit{Tuple}\\
\mathit{Rel1},\mathit{Rel2},\mathit{Rel3},\ldots <: \mathit{Relation}
\end{gather*}

We also introduce constructor functions for tuples of any arity
greater than one, for instance,
\begin{gather*}
  \mathit{binary} : \mathit{Atom} \times \mathit{Atom} \rightarrow \mathit{Tuple2}\\
  \mathit{ternary} : \mathit{Atom} \times \mathit{Atom} 
                    \times \mathit{Atom} \rightarrow \mathit{Tuple3}
\end{gather*}

We use additional axioms to specify that (1) the image of a constructor contains all tuples
of its particular arity, and (2) constructor invocations are equal iff all of
their parameters are equal.

Having these notions defined, it is straightforward
to define relational operators. A drawback of this translation approach is that
relations of different arities have to be treated separately. For every arity, a
distinct set of operators has to be defined (which we denote by subscripting the
operator names with the arities they are defined for). For example, the
cartesian product of two unary relations $r$ and $s$ is defined by
\begin{gather*}
  \mathit{prod}_{1{\times}1} : \mathit{Rel1} \times \mathit{Rel1} \rightarrow \mathit{Rel2}\\
  \forall a,b: \mathit{Atom}; (in(\mathit{binary}(a,b),\mathit{prod}_{1{\times}1}(r,s)) 
      ~\Leftrightarrow~ in(a,r) \wedge in(b,s))
\end{gather*}

Signatures and fields of an Alloy model are represented 
by constant function symbols of the appropriate type. The
address book example of Figure~\ref{example} declares four signatures and two
relations that give rise to the following declarations:
\begin{gather*}
  \mathit{Target}, \mathit{Name}, \mathit{Address}, \mathit{Book} : \mathit{Rel1}\\
  \mathit{names} : \mathit{Rel2}\qquad \mathit{addr} : \mathit{Rel3}
\end{gather*}

In order for these to only capture admissible instances of the model, we
restrict the constants to meet additional model constraints:
(1) The signatures \code{Name} and \code{Address} are disjoint subsets of \code{Target},
(2) \code{Target} is abstract, 
(3) The fields \code{names} and \code{addr} are bound by their appropriate types and
respect the multiplicity constraints.

Since every Alloy entity has a counterpart in FOL, translating an Alloy formula is
straightforward and preserves its structure. To prove a property correct, we
construct a proof obligation stating that the desired assertion follows from the
model constraints.

Alloy formulas are translated using the operators from the relational
first-order theory. By solely applying their definitions, we can rewrite the
formulas to equivalent ones, in which only the uninterpreted symbols (i.e.\ the
membership predicate $in$ and the constructor functions) appear, but none of the
operators. Although this approach might be appropriate for a purely automatic verification engine,
we consider this not suitable for the task of interactive proving since it
breaks any correspondence between the model and the proof. Moreover, this approach is 
inefficient in many cases, because formulas grow significantly in size and will
contain a lot of quantifiers. We therefore define numerous inference rules that
allow efficient reasoning on a higher abstraction level. These rules can be seen
as lemmas and have been proven to follow from the axioms of the relational
theory. Consider these representatives:
$$
  \mathsf{unionSubset}\;\frac{r \subseteq s}{~union_1(r,s) \leadsto s~}
  \qquad
  \mathsf{useSubset}\;\frac{~in(a,r)\quad r \subseteq s~}{in(a,s)}
$$

The first one is a conditional simplification rule. When the premise $r
\subseteq s$ holds, it rewrites $union_1(r,s)$ to $s$. The {\sf useSubset} rule
has two conditions, namely $in(a,r)$ and $r \subseteq s$, and infers a new formula
$in(a,s)$.

The {\key} system's proof search strategy has been adjusted to automatically apply most of the
lemmas that were defined. Although the strategy is usually not
capable of proving functionally complex properties, our tests show that the
necessary user interaction is often narrowed down to the most central steps of
the proof, and most subgoals can be closed automatically.

We illustrate our interactive reasoning approach by proving the
\code{lookupYields} assertion in the address book model:
\begin{small}
\begin{relbox}
\ul \assert lookupYields \{
\ul 	traces[] \=> (\all* b: Book, n: b.names \| \some lookup[b, n])
\ul \}.
\end{relbox}
\end{small}
The assertion states that every name known in an address book is actually
mapped to some address. This assertion is valid only if all books are
constructed by proper insertions and deletions. We therefore assume the
\code{traces} predicate to hold. Since Z3 can not prove this assertion (it times out), we
hand it to the {\key} system.

The address book model linearly orders the elements of type \code{Book} by
importing the \code{ordering} module. To reflect the linear ordering in the
translation, we define a function $b$ to enumerate all elements of $Book$, using
\key's built-in integer type $int$.  The following axioms make $b$ a bijection
from the non-negative integers to $Book$\footnote{Note that this makes $Book$ an
  infinite set. However, for this particular model, the proof remained correct
  when we changed the bijection from a function over all non-negative
  integers to one over a finite interval, thus finitizing $Book$.}:
\begin{gather*}
  b : \mathit{int} \rightarrow \mathit{Atom} \quad\quad \forall a: \mathit{Atom}; (in(a,\mathit{Book}) \Rightarrow
  \exists i: \mathit{int};(i \geq 0 \wedge a \doteq b(i)))\\
  \forall i: \mathit{int}; (i \geq 0 \Rightarrow in(b(i),\mathit{Book}))\quad\quad\forall i,j: \mathit{int}; (i,j
  \geq 0 \Rightarrow (b(i) \doteq b(j) \Leftrightarrow i \doteq j))
\end{gather*}

While proving the assertion, we face two main challenges: (1) The
\code{traces} predicate defines \code{Book} inductively. So we use induction on 
non-negative integers to prove the objective for all elements of $Book$. (2) The
\code{lookup} function uses the transitive closure of a mutable relation, namely
\code{addr}, that changes due to insertions and deletions. We therefore
use an induction principle for transitive closure, which is defined by the following
rule for an arbitrary parameterized formula $\phi$:
$$\mathsf{tc\_induct}
\frac{
  \begin{array}{l}
  \forall a,b:\mathit{Atom}; (in(\mathit{binary}(a,b),r) \Rightarrow \phi(a,b))
  \\
  \forall a,b,c: \mathit{Atom}; (in(\mathit{binary}(a,b),r)\;\wedge\\\qquad in(\mathit{binary}(b,c),\mathit{transClos}(r))
    \wedge \phi(b,c) \Rightarrow \phi(a,c)
  \end{array}
}{
  \forall a,b: Atom; (in(\mathit{binary}(a,b),\mathit{transClos}(r)) \Rightarrow \phi(a,b))
}
$$

The instantiation of $\phi$ for this rule and the induction
hypothesis have to be provided manually and require a solid insight on the
model. The remaining interactive steps provide technical guidance of the prover.
Overall, out of a total of 6522 rule applications necessary to prove \code{lookupYields}, 122 were interactive, which
included the above \textsf{tc\_induct} rule~(2 times), induction over the elements
of $Book$~(2 times), quantifier instantiations~(52), case distinctions~(17),
hiding of unnecessary formulas~(25), and miscellaneous minor steps~(24).




\section{Experiments}

In this section, we first summarize our analysis of the address book example, and then report on applying our framework to Microsoft COM standard and the mark-and-sweep garbage collection algorithm.

Table \ref{tab:addressbook} shows the performance of our framework on the address book example, and compares the results with the Alloy Analyzer (AA). The time (in seconds) is measured on an Intel Core2Quad, 2.8GHz, 8GB memory. The Alloy analysis time is the total of the time spent on generating CNF by AA 4.1.10 and solving it using the SAT4J solver. We used  Z3 version 2.19 as the underlying SMT solver. Table \ref{tab:addressbook} shows the typical progression of an integrated formal design process: First the model is analyzed using decidable, yet bounded technologies to identify errors in the specification (row 1). Having corrected the errors (row 2), the full (unbounded) verification is launched to automatically prove the properties. Since the unbounded problem is significantly harder and undecidable in general, we may have to resort to interactive verification (row 3). The steps in this row indicate the number of interactive rule applications required.

\begin{center}
\begin{table*}[tbp]
\begin{tabular}[c]{c@{ }c@{ }|c@{ }|@{ }r@{ }|@{ }c@{ }|r|c@{ }|r|@{ }c@{ }|r|}
\cline{3-10}  

& 						&\multicolumn{2}{c@{ }|@{ }}{\textsc{AA}} &  \multicolumn{2}{c@{ }|}{\textsc{bounded Z3}} & \multicolumn{2}{c@{ }|@{ }}{\textsc{unbounded Z3}} & \multicolumn{2}{c@{ }|}{\textsc{\KeY}}\\
\hline
\multicolumn{1}{|c|}{\textsc{Property}}	& \textsc{Scope}	& \textsc{Time} & \textsc{res}& \textsc{Time}& \textsc{Res} & \textsc{Time}& \textsc{Res} & \textsc{Step} & \textsc{Res}	\\
\hline                                                                                                                                                              
%
%
\multicolumn{1}{|c|}{delUndoesAdd}   			& 16 	  		& 0.8      	& CE 	       	 & 0.4	        & \cellcolor[gray]{0.8}CE	&  	& 				&    &			\\
\cline{2-6}                                                                                                                                                                  
\multicolumn{1}{|c|}{-Buggy} 				& 32 	   		& 58	  	& CE 	  	 & 1.0		& \cellcolor[gray]{0.8}CE 	& --	& NA				& -- & NA	 	\\                                                                                                                                          
\hline                                                                                                                                                                 
\hline                                                                                                                                                                 
%
%
\multicolumn{1}{|c|}{delUndoesAdd} 			& 32 	   		& 150	 	 & BV   	 & 0.0 		& BV				&  	&\cellcolor[gray]{0.8}  	&    &			\\
\cline{2-6}                                                                                                                                                        
\multicolumn{1}{|c|}{}   				& 64 	  		& TO      	 & UK		 & 0.0		& BV				& 0.0 	&\cellcolor[gray]{0.8}FV	& -- &	NA		\\
\hline                                                                                                                                                          
\hline                                                                                                                                                          
%
%
\multicolumn{1}{|c|}{lookupYields} 			& 8 	   		& 101 	 	 & BV		 & 147		& BV				& 	& 		        	&	&\cellcolor[gray]{0.8}	 \\ 
\cline{2-6}                                                                                                                                                     
\multicolumn{1}{|c|}{}					& 16 	  		& TO      	 & UK		 & TO		& UK				& TO 	& UK				& 122	&\cellcolor[gray]{0.8}FV \\
\hline
\vspace{-2mm}
 \end{tabular}
 \caption{Evaluation results for the address book example -- \textbf{CE:} counterexample, \textbf{BV:} bounded valid, \textbf{FV:} fully valid, \textbf{UK:} unknown, \textbf{NA:} not applicable, \textbf{TO:} time out ($> 10\ min.$)}
 \label{tab:addressbook}
\end{table*}
\end{center}

\begin{center}
\begin{table*}[tbp]
\begin{tabular}[c]{cc@{ }|c@{ }|r@{ }|@{ }c@{ }|r|c@{ }|r|@{ }c@{ }|r|}
\cline{3-10}  

& 						&\multicolumn{2}{c@{ }|@{ }}{\textsc{AA}} &  \multicolumn{2}{c@{ }|}{\textsc{bounded Z3}} & \multicolumn{2}{c@{ }|@{ }}{\textsc{unbounded Z3}} & \multicolumn{2}{c@{ }|}{\textsc{\KeY}}\\
\hline
\multicolumn{1}{|c|}{\textsc{Property}}	& \textsc{Scope}	& \textsc{Time} & \textsc{res}& \textsc{Time}& \textsc{Res} & \textsc{Time}& \textsc{Res} & \textsc{Time} & \textsc{Res}	\\
\hline                                                                                                                                                              
%
%
\multicolumn{1}{|c|}{BuggyCOM} 			& 16 	   		& 427	  	& CE 	  	 & 3.6		& \cellcolor[gray]{0.8}CE 	&  	& 				&    &			\\
\cline{2-6}   
\multicolumn{1}{|c|}{\emph{Theorem 1}}	 	& 17 	   		& TO	  	& CE 	  	 & 1.9		& \cellcolor[gray]{0.8}CE 	& --	& NA				& -- & NA	 	\\
\hline                                                                                                                                                                 
\hline                                                                                                                                                                 
%
%
\multicolumn{1}{|c|}{COM} 			& 16 	   		& 451	 	 & BV   	 & 0.0 		& BV				&  	&\cellcolor[gray]{0.8}  	&    &			\\
\cline{2-6}                                                                                                                                                        
\multicolumn{1}{|c|}{\emph{Theorem 1}}   	& 17 	  		& TO      	 & UK		 & 0.3		& BV				& 0.0 	&\cellcolor[gray]{0.8}FV	& -- &	NA		\\
\hline                                                                                                                                                          
\hline                                                                                                                                                          
%
%
\multicolumn{1}{|c|}{mark sweep} 		& 9 	   		& 140	 	 & BV		 & 17	  	& BV				& 	& 		        	&	&\cellcolor[gray]{0.8}	 \\ 
\cline{2-6}                                                                                                                                                     
\multicolumn{1}{|c|}{\emph{Soundness 1}}	& 10 	  		& TO      	 & UK		 & 107		& BV				& TO 	& UK				& 10	&\cellcolor[gray]{0.8}FV \\
\hline
\vspace{-2mm}
 \end{tabular}
 \caption{Evaluation results for other case studies -- \textbf{CE:} counterexample, \textbf{BV:} bounded valid, \textbf{FV:} fully valid, \textbf{UK:} unknown, \textbf{NA:} not applicable, \textbf{TO:} time out ($> 10\ min.$)}
 \label{tab:experiments}
\end{table*}
\end{center}

To further evaluate our framework, we have checked Microsoft's Component Object Model (COM) standard~\cite{com-book}, a component integration architecture that is adopted by numerous software component vendors, and provides the basis for higher-level standards such as OLE/ActiveX and COM+. We have also checked the mark-and-sweep garbage collection algorithm, which is widely used for memory management. The original Alloy specifications of these two systems are distributed with the Alloy Analyzer. However, since neither model contains an invalid assertion, we seeded a bug in the COM model (in the \emph{Identity} axiom) for a case with a counterexample (denoted by $BuggyCOM$). The results of our experiments on these two systems are given in Table \ref{tab:experiments}.

As shown in these tables, in most cases, the runtime of our bounded verification is significantly better than AA. This is because unlike AA that flattens all formulas for all possible values in the propositional form, our translation preserves the structure of the formulas, and thus exploits high-level simplifications offered by Z3. However, for \code{lookupYields}, our bounded verification does not perform as well as Alloy. This is because traces used in this assertion pose a challenge for our current translation. We are investigating further optimizations to simplify our axiomatization of traces in order to improve our performance.

Bounded verification of \code{COM-Theorem1} and \code{delUndoesAdd} shows another advantage of using QBVF. These properties have no counterexamples, and Z3 can deduce that in almost zero seconds, independent of the analyzed scope. This is because the decision procedure for QBVF can potentially deduce a contradiction from the quantified formulas independently from the sizes of the bitvectors. That is, the same quantifier instantiations that produce a contradiction for smaller bitvectors can produce contradictions for larger bitvectors without any significant overhead.

So far in our experiments, our SMT-based, unbounded verification has either proved a property correct or timed out with an inconclusive (unknown) result. A third outcome is also possible where this phase finds a counterexample that was missed by the previous bounded verification phase (because the model was not analyzed in a big-enough scope). Our current experiments, however, do not expose this case. In cases where Z3 cannot verify a property (e.g. \code{mark-and-sweep} and \code{lookupYields}), the \KeY interactive prover is invoked. As explained in Section \ref{itp}, in order to prove \code{lookupYields}, the user needs to guide \KeY by manually selecting 122 rule applications out of a total of 6522 rules necessary (the rest are selected automatically by \KeY). In the \code{mark-and-sweep} case, however, a complete  proof is found fully automatically by \KeY in 10 seconds.

\section{Conclusions}

The framework presented in this paper supports modeling, checking, and proving properties of critical infrastructures expressed in the Alloy language. It offers an economical approach by introducing a new dual-analysis engine that is capable of finding counterexamples in faulty systems and proving properties of sound systems. The analysis starts with our SMT-based, bounded, fully automatic, and decidable  verification technique that aims at finding counterexamples, and potentially outperforms AA. In this phase, the user can increase the scope in order to gain more confidence about the correctness of the property before switching to the full-verification mode. Full verification starts with our SMT-based, fully automatic proof engine, and switches to our interactive, ITP-based, complete verification only if the automatic proof engine fails.

\bibliography{biblio_02}

\end{document}